\pdfoutput=1
\RequirePackage{ifpdf}
\ifpdf 
\documentclass[pdftex]{sigma}
\else
\documentclass{sigma}
\fi

\newcommand{\pa}{\partial}

\newcommand{\no}{\nonumber}

\begin{document}

\allowdisplaybreaks

\renewcommand{\PaperNumber}{006}

\FirstPageHeading

\ShortArticleName{On the $N$-Solitons Solutions in  the Novikov--Veselov Equation}

\ArticleName{On the $\boldsymbol{N}$-Solitons Solutions\\ in  the Novikov--Veselov Equation}

\Author{Jen-Hsu CHANG}

\AuthorNameForHeading{J.H.~Chang}

\Address{Department of Computer Science and Information Engineering,\\  National Defense University,  Tauyuan, Taiwan}
\Email{\href{mailto:jhchang@ndu.edu.tw}{jhchang@ndu.edu.tw}}

\ArticleDates{Received October 01, 2012, in f\/inal form January 12, 2013; Published online January 20, 2013}

\Abstract{We construct  the $N$-solitons solution in the Novikov--Veselov
equation from the extended Moutard transformation and the Pfaf\/f\/ian
structure. Also, the corresponding wave functions are obtained
explicitly. As a result, the property characterizing the
$N$-solitons wave function is proved using the Pfaf\/f\/ian expansion.
This property corresponding to the discrete scattering data for
$N$-solitons solution is obtained in [arXiv:0912.2155] from the $\overline\partial$-dressing method.}


\Classification{35C08; 35A22}

\section{Introduction}

The Novikov--Veselov equation \cite{ba,gm,nv,vn}  is
def\/ined by
\begin{gather}
 U_t  =  \pa_z^3 U+ {\bar{\pa}}_z U+ 3\pa_z(VU)+3 {\bar{\pa}}_z
 (V^{\ast}U ), \label{NV} \\
  {\bar{\pa}}_z V = \pa_z U, \qquad
  \pa_z V^{\ast}  =  {\bar{\pa}}_z U. \nonumber
 \end{gather}
 When $z=\bar{z}=x$, we get the famous KdV equation ($U=\bar{U}=V=\bar{V}$)
 \[ U_t=2U_{xxx}+12UU_x. \]
  The equation (\ref{NV}) can be represented as the form of Manakov's triad \cite{ma}
 \[H_t=[A, H]+BH, \]
 where $H$ is the two-dimension Schr\"odinger operator
 \[
 H= \pa_z \bar{\pa_z} +U
 \]
 and
 \[
 A= \pa_z^3 + V\pa_z+ {\bar{\pa}}_z^3+\bar{V}{\bar{\pa}}_z,
 \qquad B= V_z+\bar{V}_{\bar{z}}.
 \]
 It is equivalent to the linear representation
 \begin{gather}
  H \psi=0, \qquad \pa_t \psi=A \psi. \label{rep}
  \end{gather}
 We see that the Novikov--Veselov equation (\ref{NV}) preserves a
 class of the purely potential self-adjoint operators~$H$. Here
 the pure potential means~$H$ has no external electric and
 magnetic f\/ields. The periodic inverse spectral problem for the
 two-dimensional Schr\"odinger operator~$H$ was investigated in
 terms of the Riemann surfaces with some group of involutions and
 the
 corresponding Prym $\Theta$-functions \cite{dk, gn,kr, m1, mi,
 no,sh}. On the other hand, it is  known that the Novikov--Veselov hierarchy is a
 special reduction of the two-component BKP hierarchy \cite{wz,os,kt} (and references
 therein). In~\cite{wz}, the authors showed that the Drinfeld--Sokolov
 hierarchy of D-type is a reduction of the  two-component BKP
 hierarchy using two dif\/ferent types of pseudo-dif\/ferential
 operators, which is dif\/ferent from Shiota's point of view~\cite{sh}. Also, in~\cite{m2}, it is shown that the Tzitzeica
 equation is a stationary symmetry of the Novikov--Veselov equation. Finally, it is worthwhile to notice that the Novikov--Veselov
 equation~(\ref{NV})
 is a special reduction of the Davey--Stewartson equation~\cite{ko, kg}.

 Let $H \psi= H \omega=0.$ Then via the Moutard transformation~\cite{an, mo,mo1, ni}
 \begin{gather}
  U(z, \bar{z})  \longrightarrow   \hat{U}(z, \bar{z})=U(z, \bar{z})
 +2\pa \bar{\pa} \ln \omega, \nonumber \\
 \psi  \longrightarrow   \theta=\frac{1}{\omega} \int (\psi\pa \omega-\omega\pa \psi)dz-(\psi\bar{\pa} \omega-\omega\bar{\pa}
 \psi)d \bar{z},  \label{mo}
\end{gather}
one can construct a new Schr\"odinger operator $\hat{H}= \pa_z
\bar{\pa_z} +\hat{U}$ and  $\hat {H}  \theta=0$. We remark that
the Moutard transformation~(\ref{mo}) is utilized to construct
the  $N$-solitons solutions of the Tzitzeica equation~\cite{ks}.

The extended Moutard transformation was established such
that $\hat{U}(t,z,\bar{z})$ and $\hat{V}(t,z,\bar{z})$ def\/ined by
\cite{hh, ms}
\begin{gather*} \hat{U}(t,z,\bar{z})  =  U(t,z,\bar{z})+2\pa
\bar{\pa} \ln  W(\psi, \omega),  \qquad
\hat{V} (t,z,\bar{z})  =  V(t,z,\bar{z})+2 \pa \pa \ln  W (\psi,
\omega),
\end{gather*}
where the skew product (alternating bilinear form)
$W$ is def\/ined by
 \begin{gather}
  W (\psi, \omega)  =  \int (\psi\pa
\omega-\omega\pa \psi)dz-(\psi\bar{\pa} \omega-\omega\bar{\pa}
 \psi)d \bar{z}+\big[\psi\pa^3 \omega-\omega\pa^3
 \psi+\omega\bar{\pa}^3 -\psi\bar{\pa}^3
 \omega   \nonumber \\
 \hphantom{W (\psi, \omega)  =}{}
 +
2\big(\pa^2\psi\pa\omega-\pa\psi\pa^2\omega\big)-2\big(\bar{\pa}^2\psi\bar{\pa}\omega-\bar{\pa}\psi\bar{\pa}^2\omega\big)
 +3 V(\psi\pa \omega-\omega\pa \psi)
 \nonumber \\
 \hphantom{W (\psi, \omega)  =}{}
 - 3\bar{V}(\psi\bar{\pa} \omega
 -\omega\bar{\pa}
 \psi)\big]dt , \label{ext}
 \end{gather}
will also satisfy the Novikov--Veselov equation.

 In \cite{bd,df,bf,gr}, the rational solutions and line
solitons of the Novikov--Veselov equation (\ref{NV}) are
constructed by the $\overline\partial$-dressing method. To get these kinds of
solutions, the scattering datum have to be delta-type and the
reality of~$U$ also puts some extra constraints on them. In~\cite{tt}, the singular rational
solutions are obtained using the extended Moutard transformation~(\ref{ext}); however, the non-singular rational solutions are
constructed in~\cite{jh}.

 Next, we construct Pfaf\/f\/ian-type solutions. Given any $N$
wave functions  $\psi_1, \psi_2, \psi_3,
\dots, \psi_N $ (or their linear combinations) of (\ref{rep}) for
f\/ixed potential $U(z, \bar{z}, t)$, the $N$-step extended Moutard
transformation can be obtained in the Pfaf\/f\/ian \cite{an, ni} (also see \cite{he, oh})
\begin{gather}
P(\psi_1, \psi_2, \psi_3, \dots, \psi_N)=
 \begin{cases}  \operatorname{Pf}(\psi_1, \psi_2, \psi_3, \dots, \psi_N)  & \mbox{if} \
N \  \mbox{even} , \\
\widetilde{\operatorname{Pf}}(\psi_1, \psi_2, \psi_3, \dots, \psi_N) & \mbox{if} \  N \
\mbox{odd} ,\end{cases} \nonumber\\
  \operatorname{Pf}(\psi_1, \psi_2,
\psi_3, \dots, \psi_N)  =
\sum_{\sigma}\epsilon(\sigma)W_{\sigma_1 \sigma_2}W_{\sigma_3
\sigma_4} \cdots W_{\sigma_{N-1} \sigma_N}, \label{pf1} \\
\widetilde{\operatorname{Pf}}(\psi_1, \psi_2, \psi_3, \dots, \psi_N)  =
\sum_{\sigma}\epsilon(\sigma)W_{\sigma_1 \sigma_2}W_{\sigma_3
\sigma_4} \cdots W_{\sigma_{N-2} \sigma_{N-1}}\psi_{\sigma_N},
\label{pf2}
\end{gather}
 where $W_{\sigma_{i} \sigma_j}
=W(\psi_{\sigma(i)}, \psi_{\sigma(j)})$ is def\/ined  by the skew
product~(\ref{ext}). The summations $\sigma$ in~(\ref{pf1}) and~(\ref{pf2}) run from over the permutations of
$\{1,2,3,\dots, N \}$ such that $ \sigma_1 < \sigma_2$, $\sigma_3 <
\sigma_4$, $\sigma_5 < \sigma_6$, $\dots $ and
$\sigma_1 < \sigma_3 < \sigma_5 <\sigma_7 <\cdots$,
with $\epsilon(\sigma)=1$ for the even permutations and
$\epsilon(\sigma)=-1$ for the odd permutations. Then the solution~$U$ and~$V$ can be expressed as~\cite{an}
\begin{gather*}
 U  =  U_0+ 2 \pa
\bar{\pa}
[ \ln P(\psi_1, \psi_2, \psi_3, \dots, \psi_N)],\qquad
V  =  V_0+ 2 \pa \pa [ \ln P(\psi_1, \psi_2, \psi_3, \dots,
\psi_N)], \label{pf3}
\end{gather*}
and the corresponding wave function is
\begin{gather}
\varphi= \frac{P(\psi_1, \psi_2, \psi_3, \dots, \psi_N,
\vartheta)}{P(\psi_1, \psi_2, \psi_3, \dots, \psi_N)},
\label{wave}
\end{gather}
where $\vartheta$ is an arbitrary wave function
dif\/ferent from $\psi_1, \psi_2, \psi_3, \dots,
\psi_N$.

The paper is organized as follows. In Section~\ref{section2}, we obtain
the $N$-solitons solutions using the extended Moutard transformation
and the Pfaf\/f\/ian expansion. Several examples are given.
In Section~\ref{section3}, the $N$-solitonic wave function is derived using
(\ref{wave}) and the Pfaf\/f\/ian expansion.
Section~\ref{section4} is used to prove a special property to characterize the
$N$-solitons wave function. Section~\ref{section5} is devoted to the concluding
remarks.

\section[$N$-solitons solutions]{$\boldsymbol{N}$-solitons solutions}\label{section2}

In this section, one uses successive iterations of the
extended Moutard transformation~(\ref{ext}) to construct
$N$-solitons solutions.

To obtain the $N$-solitons
solutions, we assume that $V=0$ in (\ref{NV}) and recall that $\pa
\bar{\pa}=\frac{1}{4} \triangle $.
One considers $U=-\epsilon \neq 0$, i.e.,
\begin{gather}
  \pa \bar{\pa} \varphi  =  \epsilon \varphi, \qquad
\varphi_t  =  \varphi_{zzz}+\varphi_{\bar{z}\bar{z}\bar{z}},
\label{pri}
\end{gather}
where $\epsilon$ is non-zero real constant.  The
general solution of (\ref{pri}) can be expressed as
\begin{gather}
 \varphi(z,
\bar{z}, t) = \int_{\Gamma} e^{(i\lambda) z+(i\lambda)^3
t+\frac{\epsilon}{i\lambda}\bar{z}+\frac{\epsilon^3}{(i\lambda)^3}t
} \nu(\lambda) d \lambda, \label{con}
\end{gather}
 where $\nu(\lambda)$ is
an arbitrary distribution and $\Gamma$ is an arbitrary path of
integration such that the r.h.s.\ of~(\ref{con}) is well def\/ined.

Next, using (\ref{pf1}) and (\ref{con}), one can construct
the $N$-solitons solutions. Let's take $\nu_m(\lambda)=\delta
(\lambda-p_m)$ and $ \nu_n(\lambda)=a_n \delta (\lambda-q_n)$,
where $p_m$, $a_n$, $q_n$ are complex numbers. Then one def\/ines
\begin{gather*}
\phi_m  =   \frac{\varphi (p_m)}{\sqrt{3}}= \frac{1}{\sqrt{3}}e^{F(p_m)},\qquad
\psi_n  =  a_n \frac{\varphi (q_n)}{\sqrt{3}}=\frac{a_n}{\sqrt{3}}
e^{F(q_n)}, 
\end{gather*}
where \[ F(\lambda)=(i\lambda)
z+(i\lambda)^3
t+\frac{\epsilon}{i\lambda}\bar{z}+\frac{\epsilon^3}{(i\lambda)^3}t
. \]
Then a direct calculation of the extended Moutard
transformation (\ref{ext}) can yield
\begin{gather}
 W(\phi_m,\psi_n)  = i a_n\frac{q_n-p_m}{q_n+p_m} e^{F(p_m)+F(q_n)}, \qquad
W(\phi_m,\phi_n)  = i
\frac{p_n-p_m}{p_n+p_m} e^{F(p_m)+F(p_n)}, \nonumber \\
W(\psi_m,\psi_n)  =  i a_m a_n \frac{q_n-q_m}{q_n+q_m}
e^{F(q_m)+F(q_n)}. \label{bac}
\end{gather} The $N$-solitons solutions are
def\/ined by
\[U(z, \bar{z}, t)=- \epsilon + 2 \pa
\bar{\pa} \ln \tau_N (z, \bar{z}, t), \qquad V(z, \bar{z}, t)=2 \pa
\pa \ln \tau_N (z, \bar{z}, t),\] and then
  \[ U(z, \bar{z}, t) \to -\epsilon  \qquad \mbox{as} \quad z \bar{z} \to \infty, \]
  where $t$ is f\/ixed.
The $\tau$-functions are def\/ined as follows. For simplicity, let's
denote
\begin{gather}  W(p_m, q_n)
 =  W(\phi_m,\psi_n), \! \qquad
W(p_m, p_n)  =   W(\phi_m,\phi_n) ,\! \qquad
W(q_m, q_n)  =   W(\psi_m,\psi_n), \!\!\label{si}
\end{gather} and notice
that
$F(-\lambda)= -F(\lambda)$.
The $\tau_N$ is def\/ined as
\begin{gather}
 \tau_N(z, \bar{z}, t)= \operatorname{Pf} (-p_1,
q_1, -p_2, q_2,-p_3, q_3, \dots, -p_N, q_N ), \label{tau}
\end{gather}
where
\begin{gather} (-p_m, -p_n)  =  W (-p_m, -p_n), \qquad
(-p_m, q_n)  =  W (-p_m, q_n) +\delta_{mn}, \nonumber\\
 (q_m, q_n)  =  W (q_m, q_n). \label{sii}
 \end{gather}
  To get the expansion of (\ref{tau}), we use the
following useful formula \cite{is,st}
\begin{gather}
\operatorname{Pf}(\mathbf{A}+\mathbf{B})=\sum_{r=0}^s \sum_{\alpha \in I_{2r}^m}
(-1)^{|\alpha|-r} \operatorname{Pf} (\mathbf{A}_{\alpha}) \operatorname{Pf}
(\mathbf{B}_{\alpha^c}), \label{exp}
\end{gather}
where $\mathbf{A}$ and
$\mathbf{B}$ are $m \times m $ matrices and $s= [m/2] $ is the
integer part of~$m/2$;
moreover, we denote by $\alpha^c$ the
complementary set of~$\alpha$ in the subset $\{1, 2 ,3 , \dots,
m\}$ which is arranged in increasing order, and
$|\alpha|=\alpha_1+ \alpha_2+ \cdots +\alpha_{2r}$ for
$\alpha=(\alpha_1,  \alpha_2,  \dots , \alpha_{2r})$. For the
case (\ref{tau}), one has
\begin{gather*}
 \mathbf{A}_N (z, \bar{z}, t) =
 \left[\begin{matrix} 0  & (-p_1, q_1) & (-p_1, -p_2)  &  (-p_1, q_2)  & \cdots & (-p_1, q_N)  \\
 (q_1, -p_1) & 0 & (q_1, -p_2) &  (q_1, q_2) & \cdots &  (q_1, q_N) \\
(-p_2, -p_1)  & ( -p_2, q_1) & 0  & (-p_2, q_2)  & \cdots &  (-p_2, q_N)\\
\vdots & \vdots & \vdots &  \cdots & \vdots  \\
(q_N, -p_1)& (q_N, q_1) & (q_N, -p_2) & (q_N, q_2) & \cdots & 0
\end{matrix} \right]
\end{gather*}
 and
\begin{gather*}
 \mathbf{B}_N=
 \left[\begin{matrix} 0  & 1 & 0 &  0  & \cdots & 0 \\
 -1  & 0 & 0  &  0 & \cdots &  0 \\
0 & 0 & 0  & 1  & \cdots &  0 \\
0 &0 & -1 & 0 & \cdots & 0 \\
\vdots & \vdots & \vdots &  \cdots & \vdots  \\
0 & 0 & 0  & 0   & \cdots &  1      \\
 0 & 0 & 0 & \cdots  & -1 & 0 \end{matrix}
  \right] ,
\end{gather*}
where $\mathbf{A}_N$ and $\mathbf{B}_N$ are $2N \times 2N$
matrices. Hence by (\ref{exp}) one can have the expansion of
(\ref{tau}), i.e.,
\begin{gather}
 \tau_N= 1+ \sum_{\ell=1}^{N} f_{\ell}
+\sum_{m=2}\left( \sum_{1\leq \ell_1 < \ell_2 < \cdots < \ell_m
\leq N} f_{\ell_1} f_{\ell_2} \cdots  f_{\ell_m} \prod_{1 \leq j <
k \leq m} \mathbb{P}_{\ell_j \ell_k}\right), \label{ta}
\end{gather}
where
\begin{gather*}
 f_{\ell}  =  i a_{\ell}
\frac{p_{\ell}+q_{\ell}}{q_{\ell}-p_{\ell}}
e^{F(q_{\ell})-F(p_{\ell})}, \qquad
\mathbb{P}_{\ell_j \ell_k}  =
\frac{(p_{\ell_j}-p_{\ell_k})(q_{\ell_j}-q_{\ell_k})(p_{\ell_j}+q_{\ell_k})(q_{\ell_j}+p_{\ell_k})}
{(p_{\ell_j}+p_{\ell_k})(q_{\ell_j}+q_{\ell_k})(p_{\ell_j}-q_{\ell_k})(q_{\ell_j}-p_{\ell_k})}.
\end{gather*}
 Here we have utilized the formula that if $\mathbf{C}$ is a
$2N \times 2N$  matrix  with $(i,j)$-th entry $
\frac{\alpha_i-\alpha_j}{\alpha_i+\alpha_j}$, then one has the
Schur identity~\cite{nim,os}
\begin{gather}
\operatorname{Pf}(\mathbf{C})=\prod_{1 \leq i <
j \leq 2N} \left(
\frac{\alpha_i-\alpha_j}{\alpha_i+\alpha_j}\right). \label{ca}
\end{gather}

Next, we illustrate the formula (\ref{ta}) (or \ref{tau})
with several examples.

  (1) One-soliton solution:
 \begin{gather*}
 \mathbf{A}_1 (z, \bar{z}, t)  =
 \left[\begin{matrix} 0 & (-p_1, q_1) \\
                (q_1, -p_1) & 0  \\
                \end{matrix} \right], \qquad
               \mathbf{B}_1  =
 \left[\begin{matrix}  0 & 1  \\
                -1  & 0      \\
                \end{matrix} \right]
                .
\end{gather*}
Then
\begin{gather*}
  \tau_1=\operatorname{Pf}
  (\mathbf{A}_1+\mathbf{B}_1)=1 +i
a_1\frac{q_1+p_1}{q_1-p_1}e^{F(q_1)-F(p_1)}.
\end{gather*}

(2) Two-solitons solution:
\begin{gather*}
 \mathbf{A}_2 (z, \bar{z}, t)  =
 \left[\begin{matrix} 0 & (-p_1, q_1) & (-p_1, -p_2)  &  (-p_1, q_2)    \\
(q_1, -p_1) & 0  & (q_1, -p_2) &  (q_1, q_2)  \\ (-p_2, -p_1)  &
( -p_2, q_1) & 0 & (-p_2, q_2)  \\
(q_2, -p_1)  &  ( q_2, q_1) & (q_2, -p_2) &  0  \\
\end{matrix}  \right],\\
\mathbf{B}_2  =
 \left[\begin{matrix}  0 & 1 & 0 &  0  \\
-1 & 0  & 0 &  0 \\ 0  & 0 & 0 & 1 \\
0  &  0 & -1 &  0  \\
\end{matrix} \right] .
\end{gather*}
 Then
 \begin{gather*}
 \tau_2=\operatorname{Pf}(\mathbf{A}_2+\mathbf{B}_2)
 = 1+i a_1 \frac{p_1+q_1}{q_1-p_1}e^{F(q_1)-F(p_1)}+i a_2
 \frac{p_2+q_2}{q_2-p_2}e^{F(q_2)-F(p_2)} \\
 \hphantom{\tau_2=}{}
  +   ia_1 i a_2 \frac{p_1+q_1}{q_1-p_1} \frac{p_2+q_2}{q_2-p_2}\frac{p_2-p_1}{p_2+p_1}
 \frac{q_2-q_1}{q_2+q_1}\frac{p_1+q_2}{q_2-p_1}
 \frac{p_2+q_1}{p_2-q_1}
  e^{F(q_1)-F(p_1)+F(q_2)-F(p_2)}
 \end{gather*}
 or
 \begin{gather}
 \tau_2 = 1+f_1+f_2+ \mathbb{P}_{12} f_1 f_2, \label{tau2}
\end{gather}
 where \begin{gather*}
f_1  =   i a_1 \frac{p_1+q_1}{q_1-p_1}e^{F(q_1)-F(p_1)},\qquad
f_2  =    i a_2  \frac{p_2+q_2}{q_2-p_2}e^{F(q_2)-F(p_2)}, \\
\mathbb{P}_{12}  =  \frac{p_1-p_2}{p_1+p_2}
 \frac{q_1-q_2}{q_1+q_2}\frac{p_1+q_2}{p_1-q_2}
 \frac{q_1+p_2}{q_1-p_2}.
\end{gather*}
 The $\tau_2$ soliton (\ref{tau2}) is also found in \cite{bd}
using the $\overline\partial$-dressing method.

 (3) Three-solitons solutin:
  \begin{gather*}
  \mathbf{A}_3 (z, \bar{z}, t)
   =
\left[\begin{matrix} 0 & (-p_1, q_1) & (-p_1, -p_2)  &  (-p_1, q_2)  & (-p_1,-p_3)    & (-p_1, q_3) \\
(q_1, -p_1) & 0  & (q_1, -p_2) &  (q_1, q_2) &   (q_1, -p_3)  &  (q_1, q_3)\\
(-p_2, -p_1)  & ( -p_2, q_1) & 0 & (-p_2, q_2)& (-p_2, -p_3)& (-p_2, q_3)  \\
(q_2, -p_1)  & ( q_2, q_1) & (q_2, -p_2) &  0 &(q_2, -p_3) & (q_2, q_3) \\
(-p_3, -p_1) & (-p_3, q_1) & (-p_3, -p_2 ) & (-p_3, q_2) & 0 & (-p_3, q_3) \\
(q_3, -p_1) & (q_3, q_1) & (q_3, -p_2) & (q_3,q_2) &(q_3, -p_3) &
0 \end{matrix} \right],
\\
  \mathbf{B}_3
 =
\left[\begin{matrix} 0 & 1 & 0 &  0  & 0 & 0 \\
-1 & 0  & 0 &  0 & 0  &  0 \\
0 & 0& 0 & 1& 0& 0  \\
0 & 0 & -1 &  0 &0 & 0 \\
0 & 0 & 0 & 0 & 0 & 1 \\
0 & 0 & 0 & 0 &-1 & 0 \end{matrix}
 \right].
\end{gather*}
 Then
 \begin{gather} \tau_3  =  \operatorname{Pf}
 (\mathbf{A}_3+\mathbf{B}_3) \nonumber\\
\hphantom{\tau_3}{}
 =  1+f_1+f_2+f_3+ \mathbb{P}_{12} f_1 f_2+
\mathbb{P}_{13} f_1 f_3 + \mathbb{P}_{23} f_2 f_3+\mathbb{P}_{12}
\mathbb{P}_{13}\mathbb{P}_{23}f_1 f_2 f_3, \label{tau3}
\end{gather}
where
\begin{gather*}
f_1  =   i a_1 \frac{p_1+q_1}{q_1-p_1}e^{F(q_1)-F(p_1)},\qquad
f_2  =    i a_2  \frac{p_2+q_2}{q_2-p_2}e^{F(q_2)-F(p_2)} ,\\
f_3  =    i a_3  \frac{p_3+q_3}{q_3-p_3}e^{F(q_3)-F(p_3)},\qquad
\mathbb{P}_{12}  =  \frac{p_1-p_2}{p_1+p_2}
 \frac{q_1-q_2}{q_1+q_2}\frac{p_1+q_2}{p_1-q_2}
 \frac{q_1+p_2}{q_1-p_2}, \\
\mathbb{P}_{13}  =  \frac{p_1+q_3}{p_1-q_3}
 \frac{q_1+p_3}{q_1-p_3}\frac{p_1-p_3}{p_1+p_3}
 \frac{q_1-q_3}{q_1+q_3}, \qquad
\mathbb{P}_{23}  =  \frac{p_2+q_3}{p_2-q_3}
 \frac{q_2+p_3}{q_2-p_3}\frac{p_2-p_3}{p_2+p_3}
 \frac{q_2-q_3}{q_2+q_3}.
\end{gather*}

\section{The wave functions}\label{section3}

In this section, one uses (\ref{wave}) to construct the
corresponding wave function of the $\tau$ function (\ref{ta}).

 From (\ref{wave}), one knows that the corresponding wave
function of the $N$-solitons (\ref{ta}) can be written as
 \begin{gather*}
\varphi_N=
\frac{P\big(\frac{\varphi(-p_1)}{\sqrt{3}},\frac{\varphi(q_1)}{\sqrt{3}},
\frac{\varphi(-p_2)}{\sqrt{3}}, \frac{\varphi(q_2)}{\sqrt{3}},
\dots, \frac{\varphi(-p_N)}{\sqrt{3}},
\frac{\varphi(q_N)}{\sqrt{3}},
\frac{\varphi(\lambda)}{\sqrt{3}}\big)}{\tau_N}.
\end{gather*}
 Using the
notations in (\ref{si}), (\ref{tau}) and (\ref{sii}), we can
express $\varphi_N $ as
\begin{gather} \varphi_N= \frac{P(-p_1, q_1, -p_2,
q_2, \dots, -p_N, q_N, \lambda)}{\tau_N}.
\label{wa}
\end{gather} But we
notice that
\begin{gather} (-p_m, \lambda)^{\sharp}  =   W (-p_m, \lambda),\qquad
(q_m, \lambda)^{\sharp}  =   W (q_m, \lambda),
\label{la}
\end{gather}
where $ (\cdot)^{\sharp}$ means there is no $\delta_{mn}$ here when
compared with (\ref{sii}). Now, let's compute $ P(-p_1, q_1$, $-p_2,
q_2, \dots, -p_N, q_N, \lambda) $ using~(\ref{exp}). In this
case,
\begin{gather*}
P(-p_1, q_1, -p_2, q_2, \dots, -p_N, q_N, \lambda) =\operatorname{Pf}
(M_N+Q_N),
\end{gather*}
where \[ M_N=\left[ \begin{matrix} A_N(z, \bar{z},t) &
\begin{matrix}
(-p_1, \lambda) & \frac{\varphi(-p_1)}{\sqrt{3}} \vspace{1mm}\\
(q_1, \lambda) & \frac{\varphi(q_1)}{\sqrt{3}} \\
\vdots & \vdots\\
(q_N, \lambda) & \frac{\varphi(q_N)}{\sqrt{3}} \\
\end{matrix} \vspace{1mm}\\
\begin{matrix}
(\lambda, -p_1) &(\lambda, q_1) & \cdots & (\lambda, q_N) \\
-\frac{\varphi(-p_1)}{\sqrt{3}} & -\frac{\varphi(q_1)}{\sqrt{3}} &
\cdots & -\frac{\varphi(q_N)}{\sqrt{3}}\end{matrix} &
\begin{matrix} 0 &
\frac{\varphi(\lambda)}{\sqrt{3}} \\
-\frac{\varphi(\lambda)}{\sqrt{3}} &0
\end{matrix} \end{matrix}
\right]
\]
and
\[ Q_N=
\left[ \begin{matrix} B_N & \begin{matrix}
0 & 0 \\ 0 & 0 \\
\vdots & \vdots\\
0 & 0 \\
\end{matrix} \\
\begin{matrix}
0 &0 & \cdots & 0 \\
0 & 0 & \cdots & 0 \end{matrix} &
\begin{matrix} 0 &
0 \\
0 &0\end{matrix} \end{matrix}
\right] . \]
Using (\ref{ca}), a simple
calculation yields
\begin{gather}
  \operatorname{Pf} (M_N+Q_N)  \nonumber\\
\qquad {}
 =  \phi \left [1+\sum_{\ell=1}^{N} h_{\ell}(\lambda)+
\sum_{m=2}^N \left (\sum_{{1} \leqq \ell_1 < \ell_2< \ell_3 <
\cdots <\ell_m \leqq N} h_{\ell_1}h_{\ell_2} \cdots h_{\ell_m}
\prod_{1 \leqq j <
k \leqq m} \mathbb{P}_{\ell_j \ell_k} \right) \right]  \nonumber \\
\qquad{} =  \phi \hat{\chi}_N (\lambda), \label{wa2}
\end{gather}
 where \begin{gather*}
\phi  =  \frac{\varphi(\lambda)}{\sqrt{3}},\qquad
 h_{\ell}(\lambda)  =  i a_{\ell}
 \frac{p_{\ell}+q_{\ell}}{q_{\ell} -p_{\ell}} \frac{p_{\ell}+\lambda}{p_{\ell}-\lambda}
 \frac{q_{\ell}-\lambda}{q_{\ell}+\lambda} e^{F(q_{\ell})-F(p_{\ell})} =f_{\ell}
 \frac{p_{\ell}+\lambda}{p_{\ell}-\lambda}
 \frac{q_{\ell}-\lambda}{q_{\ell}+\lambda},\\
\hat{\chi}_N (\lambda)  =  1+\sum_{\ell=1}^{N} h_{\ell}(\lambda)+
\sum_{m=2}^N \left (\sum_{{1} \leqq \ell_1 < \ell_2< \ell_3 <
\cdots <\ell_m \leqq N} h_{\ell_1}h_{\ell_2} \cdots h_{\ell_m}
\prod_{1 \leqq j < k \leqq m} \mathbb{P}_{\ell_j \ell_k} \right),
\end{gather*}
 and $\mathbb{P}_{\ell_j \ell_k}$ is def\/ined in~(\ref{ta}).

 We give several examples here.

(1) The one-soliton wave function:
\begin{gather*} M_1=\left[ \begin{matrix} A_1(z, \bar{z},t) &
\begin{matrix}
(-p_1, \lambda) & \frac{\varphi(-p_1)}{\sqrt{3}} \vspace{1mm}\\
(q_1, \lambda) & \frac{\varphi(q_1)}{\sqrt{3}} \\
\end{matrix} \vspace{1mm}\\
\begin{matrix}
(\lambda, -p_1) &(\lambda, q_1) \\
-\frac{\varphi(-p_1)}{\sqrt{3}} & -\frac{\varphi(q_1)}{\sqrt{3}}
 \end{matrix} &
\begin{matrix} 0 &
\frac{\varphi(\lambda)}{\sqrt{3}} \\
-\frac{\varphi(\lambda)}{\sqrt{3}} &0
\end{matrix} \end{matrix}\right],
\qquad
 Q_1=\left[ \begin{matrix} B_1 & \begin{matrix}
0 & 0 \\ 0 & 0 \\
\end{matrix} \\
\begin{matrix}
0 &0  \\
0 & 0 \end{matrix} &
\begin{matrix} 0 &
0 \\
0 &0 \end{matrix} \end{matrix}
\right] .
\end{gather*}
Then
\begin{gather*}
 \varphi_1  =
 \frac{P(-p_1, q_1, \lambda)}{\tau_1} =\frac{\operatorname{Pf} (M_1+Q_1)}{\tau_1}
 =  \frac{\phi}{\tau_1} \left( 1+ i a_1 \frac{p_1+q_1}{q_1 -p_1}
\frac{p_1+\lambda}{p_1-\lambda}
 \frac{q_1-\lambda}{q_1+\lambda} e^{F(q_1)-F(p_1)} \right).
\end{gather*}
 We remark that
 \begin{gather}
  \varphi_1 =  \frac{\phi}{1+f_1}\left[1+ i a_1
\frac{p_1+\lambda}{p_1-\lambda}
 \frac{q_1-\lambda}{q_1+\lambda} f_1\right]
  =  \frac{\phi}{1+f_1}\left[1+ i a_1
\left(\frac{2p_1}{p_1-\lambda}-1\right)
 \left(\frac{2 q_1}{q_1+\lambda}-1\right) f_1\right] \nonumber \\
\hphantom{\varphi_1}{}
=  \phi \left [ \frac{(1+f_1)+ 2ia_1\big(\frac{p_1}{p_1-\lambda}- \frac{q_1}{q_1+\lambda}\big)
 e^{F(q_1)-F(p_1)}} {1+f_1}\right] \nonumber \\
\hphantom{\varphi_1}{}
 = \phi \left [1-2ia_1
 \left(\frac{p_1}{\lambda-p_1}+\frac{q_1}{\lambda+q_1}\right)\frac{e^{F(q_1)-F(p_1)}}{\tau_1}\right ].
 \label{one}
\end{gather}
This is the one-soliton wave function in \cite[p.~9]{bd}.

(2) The two-soliton wave function:
\begin{gather*}
 M_2=\left[ \begin{matrix} A_2(z, \bar{z},t) &
\begin{matrix}
(-p_1, \lambda) & \frac{\varphi(-p_1)}{\sqrt{3}} \vspace{1mm}\\
(q_1, \lambda) & \frac{\varphi(q_1)}{\sqrt{3}} \vspace{1mm}\\
(-p_2, \lambda) & \frac{\varphi(-p_2)}{\sqrt{3}} \vspace{1mm}\\
(q_2, \lambda) & \frac{\varphi(q_2)}{\sqrt{3}}
\end{matrix} \vspace{1mm}\\
\begin{matrix}
(\lambda, -p_1) & (\lambda, q_1)  & (\lambda, -p_2) & (\lambda, q_2) \\
-\frac{\varphi(-p_1)}{\sqrt{3}} & -\frac{\varphi(q_1)}{\sqrt{3}} &
-\frac{\varphi(-p_2)}{\sqrt{3}} & -\frac{\varphi(q_2)}{\sqrt{3}}
 \end{matrix} &
\begin{matrix} 0 &
\frac{\varphi(\lambda)}{\sqrt{3}} \\
-\frac{\varphi(\lambda)}{\sqrt{3}} &0
\end{matrix} \end{matrix}\right],
\\
 Q_2=\left[ \begin{matrix} B_2 & \begin{matrix}
0 & 0  \\ 0 & 0 \\ 0 & 0 \\  0 & 0 \\
\end{matrix} \\
\begin{matrix}
0 &0 &0 &0  \\
0 & 0 &0 & 0 \end{matrix} &
\begin{matrix} 0 &
0 \\
0 &0 \end{matrix} \end{matrix}\right].
\end{gather*}
 Then using (\ref{wa2}), one
has \begin{gather*}
 \varphi_2  =
 \frac{P(-p_1, q_1, -p_2, q_2,  \lambda)}{\tau_2} =\frac{\operatorname{Pf}(M_2+Q_2)}{\tau_2}
 =  \frac{\phi}{\tau_2}[1+h_1(\lambda)+ h_2(\lambda)+
\mathbb{P}_{12}h_1(\lambda)h_2(\lambda) ]\\
\hphantom{\varphi_2}{}
= \frac{\phi}{\tau_2} \left( 1+ i a_1 \frac{p_1+q_1}{q_1 -p_1}
\frac{p_1+\lambda}{p_1-\lambda}
 \frac{q_1-\lambda}{q_1+\lambda} e^{F(q_1)-F(p_1)}
 +
 i a_2 \frac{p_2+q_2}{q_2 -p_2}
\frac{p_2+\lambda}{p_2-\lambda}
 \frac{q_2-\lambda}{q_2+\lambda} e^{F(q_2)-F(p_2)} \right. \\
 \left.
\hphantom{\varphi_2=}{}
   + i a_1 i a_2 \frac{p_1+q_1}{q_1 -p_1}
\frac{p_1+\lambda}{p_1-\lambda}
 \frac{q_1-\lambda}{q_1+\lambda}\frac{p_2+q_2}{q_2 -p_2}
\frac{p_2+\lambda}{p_2-\lambda}
 \frac{q_2-\lambda}{q_2+\lambda}e^{F(q_1)+F(q_2)-F(p_1)-F(p_2)}
 \right).
 \end{gather*}

 (3) Three-solitons wave function:
\begin{gather*}
 M_3=\left[ \begin{matrix} A_3(z, \bar{z},t) &
\begin{matrix}
(-p_1, \lambda) & \frac{\varphi(-p_1)}{\sqrt{3}} \vspace{1mm}\\
(q_1, \lambda) & \frac{\varphi(q_1)}{\sqrt{3}} \vspace{1mm}\\
(-p_2, \lambda) & \frac{\varphi(-p_2)}{\sqrt{3}} \vspace{1mm}\\
(q_2, \lambda) & \frac{\varphi(q_2)}{\sqrt{3}} \vspace{1mm}\\
(-p_3, \lambda) & \frac{\varphi(-p_3)}{\sqrt{3}} \vspace{1mm}\\
(q_3, \lambda) & \frac{\varphi(q_3)}{\sqrt{3}}
\end{matrix} \vspace{1mm}\\
\begin{matrix}
(\lambda, -p_1) & (\lambda, q_1)  & (\lambda, -p_2) & (\lambda,
q_2) &  (\lambda, -p_3) & (\lambda, q_3)  \\
-\frac{\varphi(-p_1)}{\sqrt{3}} & -\frac{\varphi(q_1)}{\sqrt{3}} &
-\frac{\varphi(-p_2)}{\sqrt{3}} & -\frac{\varphi(q_2)}{\sqrt{3}} &
-\frac{\varphi(-p_3)}{\sqrt{3}} & -\frac{\varphi(q_3)}{\sqrt{3}}
 \end{matrix} &
\begin{matrix} 0 &
\frac{\varphi(\lambda)}{\sqrt{3}} \\
-\frac{\varphi(\lambda)}{\sqrt{3}} &0
\end{matrix} \end{matrix}\right],
\\
 Q_3=\left[ \begin{matrix} B_3 & \begin{matrix}
0 & 0  \\ 0 & 0 \\ 0 & 0 \\  0 & 0 \\  0 & 0 \\  0 & 0
\end{matrix} \\
\begin{matrix}
0 &0 &0 &0  & 0 & 0\\
0 & 0 &0 & 0 & 0 & 0 \end{matrix} &
\begin{matrix} 0 &
0 \\
0 &0 \end{matrix} \end{matrix}\right].
\end{gather*}
From (\ref{wa2}), we get
\begin{gather*}
 \varphi_3  =
 \frac{P(-p_1, q_1, -p_2, q_2, -p_3, q_3,  \lambda)}{\tau_3} =\frac{\operatorname{Pf}(M_3+Q_3)}{\tau_3} \\
\hphantom{\varphi_3 }{}
 =  \frac{\phi}{\tau_3}[1+h_1(\lambda)+ h_2(\lambda)+h_3(\lambda)+
\mathbb{P}_{12}h_1(\lambda)h_2(\lambda)+
\mathbb{P}_{13}h_1(\lambda)h_3(\lambda)+\mathbb{P}_{23}h_2(\lambda)h_3(\lambda)\\
\hphantom{\varphi_3 =}{}
 + \mathbb{P}_{12} \mathbb{P}_{13}
\mathbb{P}_{23}h_1(\lambda)h_2(\lambda) h_3(\lambda)],
\end{gather*}
 where
\begin{gather*}
 h_1  =  i a_1 \frac{p_{1}+\lambda}{p_{1}-\lambda}
 \frac{q_{1}-\lambda}{q_{1}+\lambda}\frac{p_1+q_1}{q_1-p_1}e^{F(q_1)-F(p_1)},\qquad
h_2  =    i a_2 \frac{p_{2}+\lambda}{p_{2}-\lambda}
 \frac{q_{2}-\lambda}{q_{2}+\lambda} \frac{p_2+q_2}{q_2-p_2}e^{F(q_2)-F(p_2)}, \\
h_3  =    i a_3  \frac{p_{3}+\lambda}{p_{3}-\lambda}
 \frac{q_{3}-\lambda}{q_{3}+\lambda}\frac{p_3+q_3}{q_3-p_3}e^{F(q_3)-F(p_3)}
\end{gather*}
and $\mathbb{P}_{12}$, $\mathbb{P}_{13}$ and $\mathbb{P}_{23}$ are
def\/ined in~(\ref{tau3}).

\section[A property of $N$-solitons wave function]{A property of $\boldsymbol{N}$-solitons wave function}\label{section4}

In this section, we will express the wave function
(\ref{wa}) as another form to generalize the equa\-tion~(\ref{one})
to $N$-solitons case.

Firstly, according to the Pfaf\/f\/ian expansion in \cite{hi},
it is not dif\/f\/icult to see that
\begin{gather}
  \widetilde{\operatorname{Pf}}(b_1, b_2,
b_3, b_4, \dots, b_{2n-1}, b_{2n}, b_{2n+1}) \nonumber \\
 \qquad{} =  \sum_{m=1}^{2n+1} (-1)^{j+m} (b_j, b_m) \widetilde{\operatorname{Pf}} (b_1,
b_2, \dots, \hat{b}_j, \dots, \hat{b}_m, \dots, b_{2n},
b_{2n+1}), \label{for} \\
\quad\quad    \mbox{for} \quad j=1, 2 , \dots, 2n+1,  \nonumber
\end{gather}
where
$\hat{b}_j$ and $ \hat{b}_m $ mean these two terms are omitted.

 Secondly, noticing (\ref{bac}) and letting
$\lambda=-p_{\alpha}$ or $\lambda=q_{\alpha}$, $\alpha=1,2,
\dots, n$, we have
\begin{gather}
\widetilde{\operatorname{Pf}}(-p_1, q_1, -p_2, q_2,
\dots, -p_{\alpha},
q_{\alpha}, -p_{\alpha+1}, q_{\alpha+1}, \dots, -p_N, q_N, -p_{\alpha}) \nonumber \\
\qquad{}= \widetilde{\operatorname{Pf}}(-p_1, q_1, -p_2, q_2, \dots, -p_{\alpha},
-p_{\alpha+1}, q_{\alpha+1},
\dots, -p_N, q_N)
= \phi(-p_{\alpha}) \hat{\chi}_N (-p_{\alpha}) , \nonumber \\
 \widetilde{\operatorname{Pf}}(-p_1, q_1, -p_2, q_2, \dots, q_{\alpha-1},
-p_{\alpha}, q_{\alpha}, -p_{\alpha+1}, q_{\alpha+1}, \dots,
-p_N, q_N,
q_{\alpha}) \nonumber \\
\qquad{}= \widetilde{\operatorname{Pf}}(-p_1, q_1, -p_2, q_2, \dots,
q_{\alpha-1}, q_{\alpha}, -p_{\alpha+1}, q_{\alpha+1},
\dots, -p_N, q_N)  = a_{\alpha}\phi(q_{\alpha}) \hat{\chi}_N (q_{\alpha}).\!\!\!
\label{pfa2}
\end{gather}
 They can be seen as follows. By (\ref{for}),
one has
\begin{gather*}   \widetilde{\operatorname{Pf}}(-p_1, q_1, -p_2, q_2, \dots,
-p_{\alpha}, q_{\alpha}, -p_{\alpha+1}, q_{\alpha+1}, \dots,
-p_N, q_N,
-p_{\alpha})  \\
 =  -(q_{\alpha}, -p_1) \widetilde{\operatorname{Pf}}(q_1, -p_2, q_2, \dots,
-p_{\alpha-1}, q_{\alpha-1}, -p_{\alpha}, -p_{\alpha+1}, \dots,
-p_N, q_N, -p_{\alpha}) \\
\quad{} +  (q_{\alpha}, q_1) \widetilde{\operatorname{Pf}}(-p_1, -p_2, q_2, \dots,
-p_{\alpha-1}, q_{\alpha-1}, -p_{\alpha}, -p_{\alpha+1}, \dots,
-p_N, q_N, -p_{\alpha}) - \cdots \\
\quad{}- (q_{\alpha}, -p_{\alpha}) \widetilde{\operatorname{Pf}}(-p_1, q_1, -p_2, q_2,
\dots, -p_{\alpha-1}, q_{\alpha-1}, -p_{\alpha+1}, \dots,
-p_N, q_N, -p_{\alpha})  +  \cdots \\
 \quad{} +  (q_{\alpha}, -p_{\alpha})^{\sharp} \widetilde{\operatorname{Pf}}(-p_1, q_1,
-p_2, q_2, \dots, -p_{\alpha-1}, q_{\alpha-1}, -p_{\alpha},
-p_{\alpha+1}, \dots,-p_N, q_N) \\
 =  [-(q_{\alpha},\! -p_{\alpha})+(q_{\alpha},
-p_{\alpha})^{\sharp}] \widetilde{\operatorname{Pf}}(-p_1, q_1,\! -p_2, q_2,
\dots,\! -p_{\alpha-1}, q_{\alpha-1}, \!-p_{\alpha}, -p_{\alpha+1},
\dots,\!-p_N, q_N)\\
 =  \widetilde{\operatorname{Pf}}(-p_1, q_1, -p_2, q_2, \dots, -p_{\alpha-1},
q_{\alpha-1}, -p_{\alpha}, -p_{\alpha+1}, q_{\alpha+1}, \dots,
-p_N, q_N) ,
\end{gather*}
 where $(q_{\alpha}, -p_{\alpha})^{\sharp}$ is
def\/ined in (\ref{la}) and we know that
\[  \widetilde{\operatorname{Pf}}(\dots,  -p_{\alpha}, \dots,-p_{\alpha})=0. \]
The second equation of (\ref{pfa2}) can be proved similarly.

Finally, from (\ref{for}), one yields
\begin{gather*}
  \widetilde{\operatorname{Pf}}(-p_1, q_1, -p_2, q_2, \dots, -p_N, q_N, \lambda)
 =  (\lambda, -p_1)\widetilde{\operatorname{Pf}}( q_1, -p_2, q_2, \dots, -p_N,
q_N)  \no \\
 \quad{} -  (\lambda, q_1)\widetilde{\operatorname{Pf}}( -p_1, -p_2, q_2, \dots, -p_N,
q_N)
 +  (\lambda, -p_2)\widetilde{\operatorname{Pf}}( q_1, -p_1, q_2, \dots, -p_N,
q_N)     \\
\quad{} -  (\lambda, q_2)\widetilde{\operatorname{Pf}}( q_1, -p_1, -p_2, \dots, -p_N,
q_N)    +  \cdots   \\
\quad{} +  (\lambda, -p_N)\widetilde{\operatorname{Pf}}( -p_1, -p_2, q_2, \dots,
-p_{N-1},
q_{N-1}, q_N)    \\
\quad{} -  (\lambda, q_N)\widetilde{\operatorname{Pf}}( -p_1, -p_2, q_2, \dots,
-p_{N-1}, q_{N-1}, -p_N) + \phi(\lambda) \tau_N. 
\end{gather*}
Also, the wave function (\ref{wa2}) can be written as
\[
\varphi_N (\lambda) =\phi (\lambda)    \chi_N (\lambda) ,
\]
where $ \chi_N (\lambda)=\frac{\hat{\chi}_N(\lambda) }{\tau_N}$.
Therefore, using (\ref{pfa2}) and letting $\Delta F_n=F(q_n)-F(p_n)$, we get
\begin{gather}
  \chi_N (\lambda)  =
1-\frac{p_1+\lambda}{\lambda-p_1} ia_1 e^{\Delta F_1} \chi_N
(q_1)-\frac{q_1-\lambda}{q_1+\lambda} ia_1 e^{\Delta F_1}
\chi_N (-p_1) \nonumber \\
\hphantom{\chi_N (\lambda)  =}{}  - \frac{p_2+\lambda}{\lambda-p_2} ia_2 e^{\Delta F_2} \chi_N
(q_2)-\frac{q_2-\lambda}{q_2+\lambda} ia_2 e^{\Delta F_2}
\chi_N (-p_2) -\cdots \nonumber \\
 \hphantom{\chi_N (\lambda)  =}{} -  \frac{p_N+\lambda}{\lambda-p_N} ia_N e^{\Delta F_N} \chi_N
(q_N)-\frac{q_N-\lambda}{q_N+\lambda} ia_N e^{\Delta F_N}
\chi_N (-p_N) \nonumber \\
\hphantom{\chi_N (\lambda)}{} =  1-\frac{2 p_1}{\lambda-p_1} ia_1 e^{\Delta F_1} \chi_N
(q_1)-\frac{2 q_1}{q_1+\lambda} ia_1 e^{\Delta F_1}
\chi_N (-p_1) \nonumber \\
\hphantom{\chi_N (\lambda)  =}{} - \frac{2 p_2}{\lambda-p_2} ia_2 e^{\Delta F_2} \chi_N
(q_2)-\frac{2 q_2}{q_2+\lambda} ia_2 e^{\Delta F_2}
\chi_N (-p_2)  - \cdots \nonumber \\
\hphantom{\chi_N (\lambda)  =}{} -  \frac{2 p_N}{\lambda-p_N} ia_N e^{\Delta F_N} \chi_N
(q_N)-\frac{2 q_N}{q_N+\lambda} ia_N e^{\Delta F_N}
\chi_N (-p_N) \nonumber \\
\hphantom{\chi_N (\lambda)  =}{} +  \big[{-} ia_1 e^{\Delta F_1} \chi_N (q_1)+ ia_1 e^{\Delta F_1}
\chi_N (-p_1) - ia_2 e^{\Delta F_2} \chi_N (q_2) \nonumber\\
\hphantom{\chi_N (\lambda)  =}{} +  ia_2 e^{\Delta F_2} \chi_N (-p_2)- \cdots -ia_N e^{\Delta F_N}
\chi_N (q_N)+ ia_N e^{\Delta F_N} \chi_N (-p_N) \big]   .\label{fin}
\end{gather}  Since
$\widetilde{\operatorname{Pf}}(-p_1, q_1, -p_2, q_2, \dots, -p_N, q_N,
0)=\frac{\tau_N}{\sqrt {3}}$, we have $\hat{\chi}_N (0)=\tau_N$.
Then the last term in $[\cdots]$ of (\ref{fin}) is zero (or
$\lim\limits_{\lambda \to \infty} \chi_N (\lambda)=1 $ ). Hence one has
\begin{gather*}
\chi_N(\lambda) =  1-\frac{2 p_1}{\lambda-p_1} ia_1 e^{\Delta
F_1} \chi_N (q_1)-\frac{2 q_1}{q_1+\lambda} ia_1 e^{\Delta F_1}
\chi_N (-p_1)   \\
\hphantom{\chi_N(\lambda) =}{}
 - \frac{2 p_2}{\lambda-p_2} ia_2 e^{\Delta F_2} \chi_N
(q_2)-\frac{2 q_2}{q_2+\lambda} ia_2 e^{\Delta F_2}
\chi_N (-p_2) -  \cdots   \\
\hphantom{\chi_N(\lambda) =}{}
 -  \frac{2 p_N}{\lambda-p_N} ia_N e^{\Delta F_N} \chi_N
(q_N)-\frac{2 q_N}{q_N+\lambda} ia_N e^{\Delta F_N} \chi_N (-p_N).
\end{gather*}
This formula is also obtained by the d-bar
dressing method when the d-bar data is the degenerate delta kernel
\cite[p.~6]{bd}.

When $n$=1, we have (\ref{one}). For  $n$=2, from
(\ref{wa2}), one knows that
\begin{gather*}
 \chi_2(-p_1)  =  \left [1+ ia_2
\frac{p_2+q_2}{q_2-p_2}
 \frac{p_2-p_1}{p_2+p_1} \frac{q_2+p_1}{q_2-p_1}e^{\triangle F_2}\right ]
 / \tau_2, \\
\chi_2(q_1)  =  \left [1+ ia_2 \frac{p_2+q_2}{q_2-p_2}
 \frac{p_2+q_1}{p_2-q_1}
 \frac{q_2-q_1}{q_2+q_1}e^{\triangle F_2}\right]/
 \tau_2, \\
 \chi_2(-p_2)  =  \left[1+ ia_1 \frac{p_1+q_1}{q_1-p_1}
 \frac{p_1-p_2}{p_1+p_2}
 \frac{q_1+p_2}{q_1-p_2}e^{\triangle F_1}\right]/
 \tau_2, \\
 \chi_2(q_2)  =  \left[1+ ia_1 \frac{p_1+q_1}{q_1-p_1}
 \frac{p_1+q_2}{p_1-q_2}
 \frac{q_1-q_2}{q_1+q_2}e^{\triangle F_1}\right]/
 \tau_2.
\end{gather*}
Then
\begin{gather*}
 \chi_2(\lambda) =  1-\frac{2 p_1}{\lambda-p_1}
ia_1 e^{\Delta F_1} \chi_2 (q_1)-\frac{2 q_1}{q_1+\lambda} ia_1
e^{\Delta F_1}
\chi_2 (-p_1)   \\
\hphantom{\chi_2(\lambda) =}{}
- \frac{2 p_2}{\lambda-p_2} ia_2 e^{\Delta F_2} \chi_2
(q_2)-\frac{2 q_2}{q_2+\lambda} ia_2 e^{\Delta F_2} \chi_2 (-p_2).
\end{gather*}
We remark that this formula also appears in \cite[p.~10]{bd}, the parameters being dif\/ferent. For $n$=3, by~(\ref{wa2}), we
obtain \begin{gather*}
 \chi_3(-p_1)  =  \left [1+ ia_2
\frac{p_2+q_2}{q_2-p_2}
 \frac{p_2-p_1}{p_2+p_1} \frac{q_2+p_1}{q_2-p_1}e^{\triangle F_2}+
 ia_3 \frac{p_3+q_3}{q_3-p_3}
 \frac{p_3-p_1}{p_3+p_1} \frac{q_3+p_1}{q_3-p_1}e^{\triangle F_3}
 \right. \\
\left.
\hphantom{\chi_3(-p_1)  =}{}
+ i a_2 i a_3\frac{p_2+q_2}{q_2-p_2}
 \frac{p_2-p_1}{p_2+p_1} \frac{q_2+p_1}{q_2-p_1}\frac{p_3+q_3}{q_3-p_3}
 \frac{p_3-p_1}{p_3+p_1} \frac{q_3+p_1}{q_3-p_1}\mathbb{P}_{23}e^{\triangle F_2 +\triangle
 F_3}\right]/\tau_3, \\
 \chi_3 (q_1)  =  \left [1+ ia_2 \frac{p_2+q_2}{q_2-p_2}
 \frac{p_2+q_1}{p_2-q_1}\frac{q_2-q_1}{q_2+q_1}e^{\triangle F_2} + ia_3 \frac{p_3+q_3}{q_3-p_3}
 \frac{p_2+q_1}{p_2-q_1} \frac{q_3-q_1}{q_3+q_1}e^{\triangle F_3} \right. \\
\left.
\hphantom{\chi_3 (q_1)  =}{}
+ ia_2 ia_3 \frac{p_2+q_2}{q_2-p_2}
 \frac{p_2+q_1}{p_2-q_1}\frac{q_2-q_1}{q_2+q_1}\frac{p_3+q_3}{q_3-p_3}
 \frac{p_2+q_1}{p_2-q_1} \frac{q_3-q_1}{q_3+q_1}\mathbb{P}_{23}e^{\triangle F_2+\triangle
 F_3} \right]/\tau_3, \\
 \chi_3(-p_2)  =  \left[1+ ia_1 \frac{p_1+q_1}{q_1-p_1}
 \frac{p_1-p_2}{p_1+p_2}
 \frac{q_1+p_2}{q_1-p_2}e^{\triangle F_1}+  ia_3 \frac{p_3+q_3}{p_3-q_3}
 \frac{p_3-p_2}{p_3+p_2} \frac{q_3+p_2}{q_3-p_2}e^{\triangle
 F_3} \right. \\
\left.
\hphantom{\chi_3(-p_2)  =}{}
+  ia_1 ia_3 \frac{p_1+q_1}{q_1-p_1}
 \frac{p_1-p_2}{p_1+p_2}
 \frac{q_1+p_2}{q_1-p_2}\frac{p_3+q_3}{p_3-q_3}
 \frac{p_3-p_2}{p_3+p_2} \frac{q_3+p_2}{q_3-p_2}\mathbb{P}_{13}e^{\triangle F_1 +\triangle
 F_3} \right]/\tau_3, \\
\chi_3(q_2)  =  \left[1+ ia_1 \frac{p_1+q_1}{q_1-p_1}
 \frac{p_1+q_2}{p_1-q_2}
 \frac{q_1-q_2}{q_1+q_2}e^{\triangle F_1}+ ia_3 \frac{p_3+q_3}{q_3-p_3}
 \frac{p_3+q_2}{p_3-q_2} \frac{q_3-q_2}{q_3+q_2}e^{\triangle
 F_3} \right. \\
\left.
\hphantom{\chi_3(q_2)  =}{}
 +  ia_1 ia_3 \frac{p_1+q_1}{q_1-p_1}
 \frac{p_1+q_2}{p_1-q_2}
 \frac{q_1-q_2}{q_1+q_2}\frac{p_3+q_3}{q_3-p_3}
 \frac{p_3+q_2}{p_3-q_2} \frac{q_3-q_2}{q_3+q_2}\mathbb{P}_{13}e^{\triangle F_1 +\triangle
 F_3} \right]/\tau_3, \\
\chi_3(-p_3)  =  \left[1+ ia_1 \frac{p_1+q_1}{p_1-q_1}
 \frac{p_1-p_3}{p_1+p_3}
 \frac{q_1+p_3}{q_1-p_3}e^{\triangle F_1}+  ia_2 \frac{p_2+q_2}{q_2-p_2}
 \frac{p_2-p_3}{p_2+p_3} \frac{q_2+p_3}{q_2-p_3}e^{\triangle
 F_2} \right. \\
\left.
\hphantom{\chi_3(-p_3)  =}{}
 +  ia_1 ia_2 \frac{p_1+q_1}{p_1-q_1}
 \frac{p_1-p_3}{p_1+p_3}
 \frac{q_1+p_3}{q_1-p_3}\frac{p_2+q_2}{q_2-p_2}
 \frac{p_2-p_3}{p_2+p_3} \frac{q_2+p_3}{q_2-p_3}\mathbb{P}_{12} e^{\triangle F_1+\triangle F_2}
\right]/\tau_3, \\
\chi_3(q_3)  =  \left[1+ ia_1 \frac{p_1+q_1}{q_1-p_1}
 \frac{p_1+q_3}{p_1-q_3}
 \frac{q_1-q_3}{q_1+q_3}e^{\triangle F_1}+ ia_2 \frac{p_2+q_2}{q_2-p_2}
 \frac{p_2+q_3}{p_2-q_3} \frac{q_2-q_3}{q_2+q_3}e^{\triangle
 F_2} \right. \\
 \left.
 \hphantom{\chi_3(q_3)  =}{}
 +  ia_1 ia_2  \frac{p_1+q_1}{q_1-p_1}
 \frac{p_1+q_3}{p_1-q_3}
 \frac{q_1-q_3}{q_1+q_3}\frac{p_2+q_2}{q_2-p_2}
 \frac{p_2+q_3}{p_2-q_3} \frac{q_2-q_3}{q_2+q_3}\mathbb{P}_{12}e^{\triangle
 F_1+\triangle F_2} \right]/\tau_3,
\end{gather*}
where $\mathbb{P}_{12}$, $\mathbb{P}_{13}$ and
$\mathbb{P}_{23}$ are def\/ined in (\ref{tau3}).  Then
\begin{gather*}
\chi_3(\lambda)  =  1-\frac{2 p_1}{\lambda-p_1} ia_1 e^{\Delta
F_1} \chi_3 (q_1)-\frac{2 q_1}{q_1+\lambda} ia_1 e^{\Delta F_1}
\chi_3 (-p_1)  -\frac{2 p_2}{\lambda-p_2} ia_2 e^{\Delta F_2} \chi_3
(q_2)\\
\hphantom{\chi_3(\lambda)  =}{}
-\frac{2 q_2}{q_2+\lambda} ia_2 e^{\Delta F_2} \chi_3 (-p_2)
- \frac{2 p_3}{\lambda-p_3} ia_3 e^{\Delta F_3} \chi_3
(q_3)-\frac{2 q_3}{q_3+\lambda} ia_3 e^{\Delta F_3} \chi_3 (-p_3).
\end{gather*}

\section{Concluding remarks}\label{section5}

In this paper, we have used the extended Moutard transformation to construct the $N$-solitons solutions. The basic
idea comes from the successive iterations of solitons solutions,
as remains to be the simple method  to obtain the $N$-solitons
solutions. Also, the corresponding wave functions are  constructed
by the Pfaf\/f\/ian expansion of the sum of two anti-symmetric
matrices (\ref{exp}) when compared
with the $\overline\partial$-dressing method \cite{bd}.

To obtain real $N$-solitons solutions of  the
Novikov--Veselov equation (\ref{NV}), one has to put extra
relations between $-p_i$ and $ q_i$ \cite{bd}. It could be
interesting to investigate these real solutions for the
Schr\"odinger operator (self-adjoint). On the other hand, the
resonance of $N$-solitons solutions of DKP or KP theory has been
studied in \cite{ko1,km, ko3, ko2}. And then the resonance of
$N$-solitons solutions of Pfaf\/f\/ian type (\ref{ta})  deserves to be
investigated. These issues will be published elsewhere.

\subsection*{Acknowledgements}

This work is supported in part by the National Science Council of
Taiwan under Grant No. 100-2115-M-606-001.

\pdfbookmark[1]{References}{ref}
\LastPageEnding

\end{document}